\documentclass[page-classic]{epl2}

\usepackage{amsmath}
\usepackage{hyperref}

\title{Effect of tax dynamics on linearly growing processes under stochastic resetting: a possible economic model}
\shorttitle{Effect of tax dynamics on linearly growing processes under stochastic resetting} %Insert here a short version of the title if it exceeds 70 characters

\author{Ion Santra}

\institute{Raman Research Institute, Bengaluru 560080, India        }

\abstract{We study a system of $N$ agents, whose wealth grows linearly, under the effect of
stochastic resetting and interacting via a tax-like dynamics—all agents donate a part of their
wealth, which is, in turn, redistributed equally among all others. This mimics a socio-economic
scenario where people have fixed incomes, suffer individual economic setbacks, and pay taxes to
the state. The system always reaches a stationary state, which shows a trivial exponential wealth
distribution in the absence of tax dynamics. The introduction of the tax dynamics leads to several interesting features in the stationary wealth distribution. In particular, we analytically
find that an increase in taxation for a homogeneous system (where all agents are alike) results in a
transition from a society where agents are most likely poor to another where rich agents are more
common. We also study inhomogeneous systems, where the growth rates of the agents are chosen
from a distribution, and the taxation is proportional to the individual growth rates. We find an
optimal taxation, which produces a complete economic equality (average wealth is independent
of the individual growth rates), beyond which there is a reverse disparity, where agents with low
growth rates are more likely to be rich. We consider three income distributions observed in the real
world and show that they exhibit the same qualitative features. Our analytical results are in
the $N\to\infty$ limit and backed by numerical simulations.}
\begin{document}

\maketitle

\section{Introduction}
Growth is one of the most interesting and omnipresent realities of our lives. Though forms of growth can be quite complex in general, possibly the simplest models of growth---linear growth is ubiquitous. Because of its simplicity, it is easier to understand it conceptually, as well as analytically.  
 The growths, however, as is a common experience for all, are not always everlasting. In economies, for example, there are catastrophic events resulting in great setbacks~\cite{economycrash1}. Mathematically such catastrophic events can be modeled effectively by stochastic resetting. In stochastic resetting, a dynamical process intermittently stops, resets to some specific value (or distribution) and then resumes again~\cite{majumdar1,res_rev}. The last decade has seen a surge of interest in the field of stochastic resetting, with its effect being studied on diffusion~\cite{majumdar1} and other diffusion like systems~\cite{difflike1,difflike2,difflike3,difflike4}, fluctuating interfaces~\cite{interface1,interface2}, interacting particle systems~\cite{interacting1,interacting2,interacting3}, active particles~\cite{biologicalsystem1,biologicalsystem2,biologicalsystem3,
 biologicalsystem4}, reaction diffusion systems, biological and chemical reactions~\cite{chemreaction1,chemreaction2,chemreaction3}, economic models of income dynamics\cite{palincome}.

In this paper, we study $N$ agents whose individual wealth grows linearly with time, along with stochastic resets at a constant rate. In addition, the agents also follow a  tax dynamics---each of them donates a part of their wealth, which is in turn redistributed among all. This is actually a very simplistic model of our society, containing some basic attributes of a real economy where working people have fixed monthly incomes and pay taxes. The resetting events, on the other hand, represent the catastrophic events where an individual suffers an economic setback. Note that the catastrophic events here do not correspond to a global catastrophe like recession or pandemic but a catastrophe at an individual level like an accident, serious illness, burglary, failed investment, etc.

Agent-based models have been successfully used to understand the wealth dynamics of a society in the past~\cite{Boghosian2019_sciam,pairwise0,pairwise1}. In most of these studies, the economic activity is considered to be a direct pairwise wealth exchange between agents, modeled by a scattering process~\cite{Slanina2004,agentbasedmodel1}. The long time behavior of the wealth distribution in these models depend on the microscopic structure of the scattering events, which are modeled to incorporate the effects of random wealth transfers~\cite{Bisi2009}, savings~\cite{Chakraborti2000,Chatterjee2004}, risky markets and stock exchanges~\cite{Cordier2009,Cordier2005}, and taxations. In contrast to these studies, our model follows a different approach where the agents do not interact directly but `indirectly' by some wealth tax dynamics~\cite{Garibaldi2007,Pianegonda2003}. It is important to note that though simple, this model shows some interesting, physically sound results and has the scope of systematic inclusion of several real world complexities.

The resetting events force the system to reach a non-equilibrium stationary state at sufficiently long times, which we study analytically. We show that in the absence of tax dynamics, the wealth distribution of each agent reaches a stationary state which is exponentially decaying. The introduction of taxation (interaction) changes this remarkably: For a homogeneous society, where all the agents are exactly alike, the wealth distribution shows a power-law behavior with finite upper and lower cut-offs. Furthermore, the wealth distribution of an agent exhibits interesting shape transitions with the strength of the interaction---at low taxation, agents are mostly poor, while at large taxation, the agents are more likely to be rich. We then move to a more realistic inhomogeneous society where each agent has a different growth rate. In this case, we show that, with all other parameters remaining homogeneous, `the rich are always richer' i.e., the stationary value of the average wealth of an agent is linearly proportional to its respective growth rate. However, if the tax paid by each agent is proportional to their respective growth rates, then this unprecedented growth can be prevented, and the average wealth always saturates to some fixed value. In fact, we find an optimal taxation at which there is complete economic equality, beyond which there is a reverse disparity where the agents with low incomes tend to have higher average wealth. To showcase the robustness of this result, we choose three different relevant growth-rate (income) distributions seen in real world economies---exponential distribution, Gamma distribution, and power-law distribution. All of our predictions are based on exact results obtained from a mean field-like approximation for the interaction and backed by numerical simulations of the exact system. 
It is important to put the early disclaimer that this is a very simplistic model of the economy of our society, not in its full complexity, and the results should be interpreted in that light only.

\begin{figure}
\includegraphics[width=0.95\hsize]{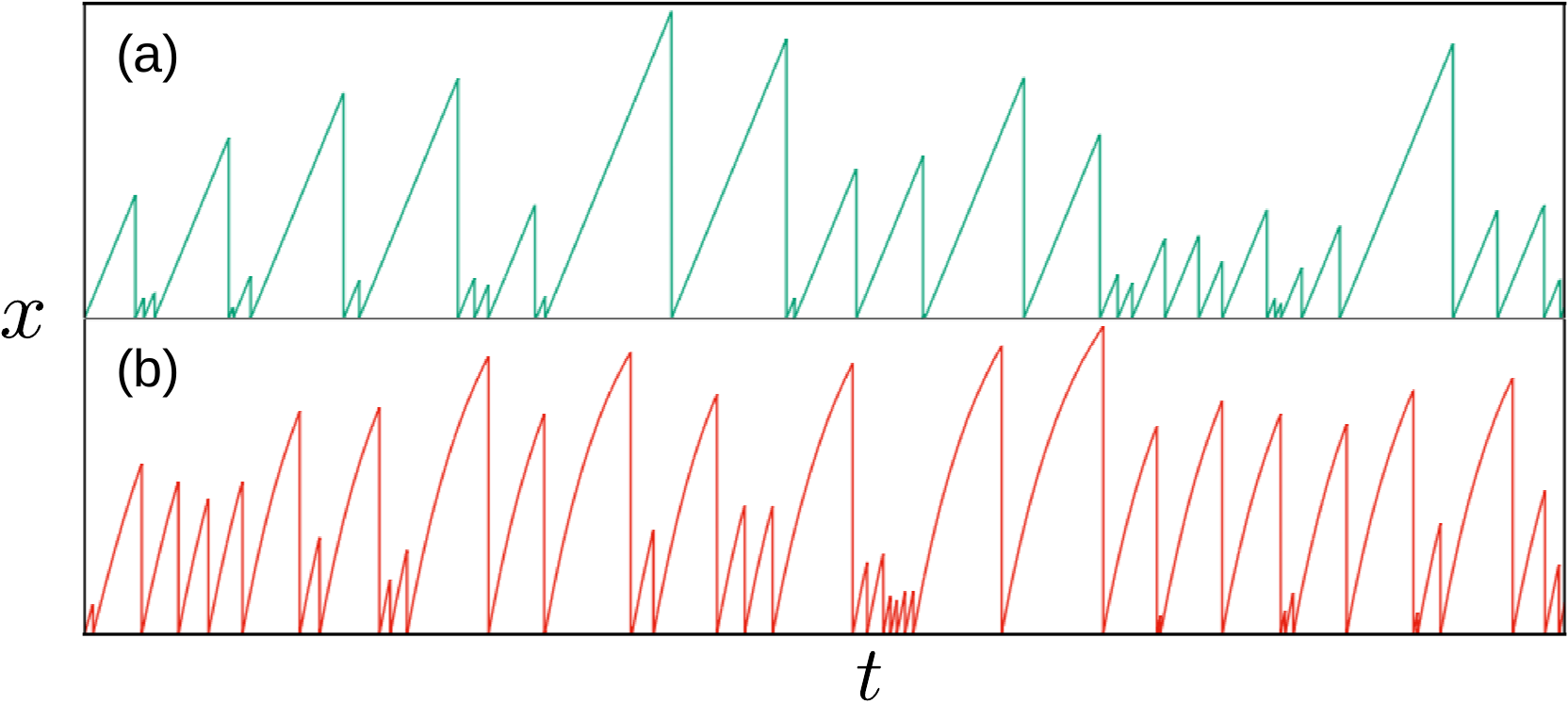}
\caption{Typical trajectory of wealth dynamics of an agent (a) in absence of tax dynamics following Eq.~\eqref{eq:L_nonint}, (b) in the presence of tax dynamics following Eq.~\eqref{eq:L_int}. In the presence of tax dynamics, panel (b), we see that between two resetting events the curve bends away from the straight line, showing an effective slow down of the growth with time. This is an indication of the upper cut-off for the distribution $P^s(x)$ at $x^*$.}
\label{f:traj}
\end{figure}
\section{Model}
 The \emph{wealth} of an agent, denoted by  $x_i$ increases at a constant \emph{growth rate} $v_i>0$ and resets stochastically at a rate $r_i$ to a value $x_r$. This is equivalent to the integrated Langevin equation,
\begin{align}
\label{eq:L_nonint}
x_i&(t+dt)=\begin{cases}x_r\quad \text{with probability }r_i dt \\
x_i(t)+v_idt\quad \text{with probability }(1-r_idt)
\end{cases}
\end{align}
where $p$ denotes the probability of the corresponding event. Note that, this dynamics is similar to a continuum version of the Sisyphus random walk~\cite{Montero2016,Montero2021}. We consider $N$ such agents, which in addition follow a tax dynamics---all the agents pay a part of their wealth $c_i x_i$, where $c_i>0$ is the \emph{collection parameter} of the $i$th agent to the government, and the government, in turn, redistributes the total collection $C=\sum_{i=1}^Nc_i x_i$ equally among all agents, with the $i$th agent receiving an amount $d_i C$ (where $d_i$ is the \emph{distribution parameter} for the $i$th agent). Thus, the full Langevin equation for the $i$th agent can be written as,
\begin{align}
x_i(t+dt)=\begin{cases}x_r\quad\text{with probability }r_idt\\
x(t)+(v_i-c_ix_i+ d_i C)dt~\text{with }\\
\qquad\text{probability }(1-r_idt). \label{eq:L_int}
\end{cases}
\end{align}
To understand the dynamics, it is instructive to first look at the distribution of wealth $P(x_i,t)$ of a single agent in the absence of the tax dynamics i.e., $c_i=0\,\forall\, i$.
\section{Wealth distribution in the absence of tax dynamics}
The Fokker-Planck equation governing the wealth distribution of an agent in the absence of tax dynamics, corresponding to the non-interacting Langevin equation \eqref{eq:L_nonint}, is given by,
\begin{align}
\frac{\partial P(x,t)}{\partial t}=-v\frac{\partial P(x,t)}{\partial x}-r P(x,t)+r\delta(x-x_r).
\end{align}
The solution of the above equation can be obtained very easily, given an initial condition, and at long-times it takes the stationary form,
\begin{align}
P(x,t\to\infty)=\frac{r}{v}\exp\left[-\frac{r}{v}(x-x_r)\right]\Theta(x-x_r),
\label{nonint}
\end{align}
where $\Theta(z)$ denotes the Heaviside theta function.
Thus the stationary wealth distribution decays exponentially at long-times. The average wealth of an agent $\bar{x}$ in the stationary state can be readily calculated and comes out to be $x_r+v/r$. Having got an idea of how the underlying resetting dynamics affect the growth, we are now in a position to study the effects of the tax dynamics.

\section{Homogeneous system with tax dynamics}
To begin with, we consider the simplest possible case---a homogeneous system of N agents where the growth rate, resetting rate, collection and distribution parameters are same for all the agents, i.e., $v_i=v,\,r_i=r,\, c_i=c, d_i=d$ for all $i$.

This automatically fixes the redistribution parameter,  $\sum_{i=1}^N d=1\implies d=N^{-1}$. The redistribution term ($R_s=d\sum_{i=1}^N cx_i$) thus becomes,
\begin{align}
\frac{R_s}{c}=d\sum_{i=1}^Nx_i=\frac{1}{N}\sum_{i=1}^Nx_i=\bar{x},
\end{align}
where $\bar{x}$ is the arithmetic mean of the wealth of all agents.
We know, from our study of the non-interacting agents that the system at large times reaches a stationary state. Thus, the average wealth of each agent also reaches a time independent stationary value. Thus, $\bar{x}$ also becomes independent of time. At this point, we make an assumption that for large system size $N\to\infty$, $\bar{x}$ should be same as the stationary state ensemble average of a single process $\langle x\rangle=\int_0^{\infty}dx\,x\, P^s(x)$, to the leading order.  This ergodicity hypothesis---that the stationary state attained due to stochastic resetting renders ergodicity, has been recently shown in \cite{ergodicity1} for diffusive systems under stochastic resetting. 

This can be justified in the following way. The fluctuation in sum of $N$ random variables is O($N^{1/2}$), thus, corresponding contribution to $R_s\sim\text{O}(N^{-1/2})$, which goes to zero for large $N$, i.e.,
\begin{align}
\bar{x}\simeq\langle x\rangle +O(N^{-\frac{1}{2}})=\langle  x\rangle  \textrm{ as }N\to\infty.
\label{ergodicity}
\end{align}
Following this assumption, at large times, we can decouple the Langevin eq.~\eqref{eq:L_int},
\begin{align}
x(t+dt)&=\begin{cases}x_r\quad \text{with probability }r_idt \\
x(t)+v_{\text{mf}}~dt~\, \text{with probability }(1-r_idt),
\end{cases}
\label{eq:L_int_mf}
\end{align}
with the net `mean-field' drift $v_{\text{mf}}=(v+c\bar{x}-c x)$. Note that, we have dropped the agent index $i$ for this section as all agents are exactly alike. The corresponding stationary Fokker-Planck equation governing the wealth distribution of an agent $P^s(x)$ can be immediately written down, 
\begin{align}
 \frac{\partial}{\partial x}[v_{\text{mf}}(x)P^s(x)]+r P^s(x)=r\delta(x-x_r).
 \label{mf:FP}
 \end{align}
The ergodicity hypothesis Eq.~\eqref{ergodicity} implies that $\bar{x}$ should satisfy the self consistency relation,
 \begin{align}
 \bar{x}=\int_0^{\infty}dx\,x\, P^s(x),
 \label{eq:selfconsist}
 \end{align}
 where $P^s(x)$ is the solution of stationary distribution which satisfies the Fokker-Planck equation Eq.~\eqref{mf:FP}.

 Equation \eqref{mf:FP} can be easily solved by solving the homogeneous equation for $x\neq x_r$ and then exploiting the discontinuity in $P^s(x)$ at $x=x_r$. In fact, $P^s(x)=0$ for $x<x_r$, since at large times (there have been atleast one reset), the wealth of any agent cannot be less than $x_r$. Using the above conditions, we obtain,
\begin{align}
P^s(x)=\frac{r}{(v+c\bar{x}-cx_r)^{r/c}}(v+c\bar{x}-cx)^{r/c-1}.
\label{eq:ss}
\end{align}
\begin{figure}
\includegraphics[width=\hsize]{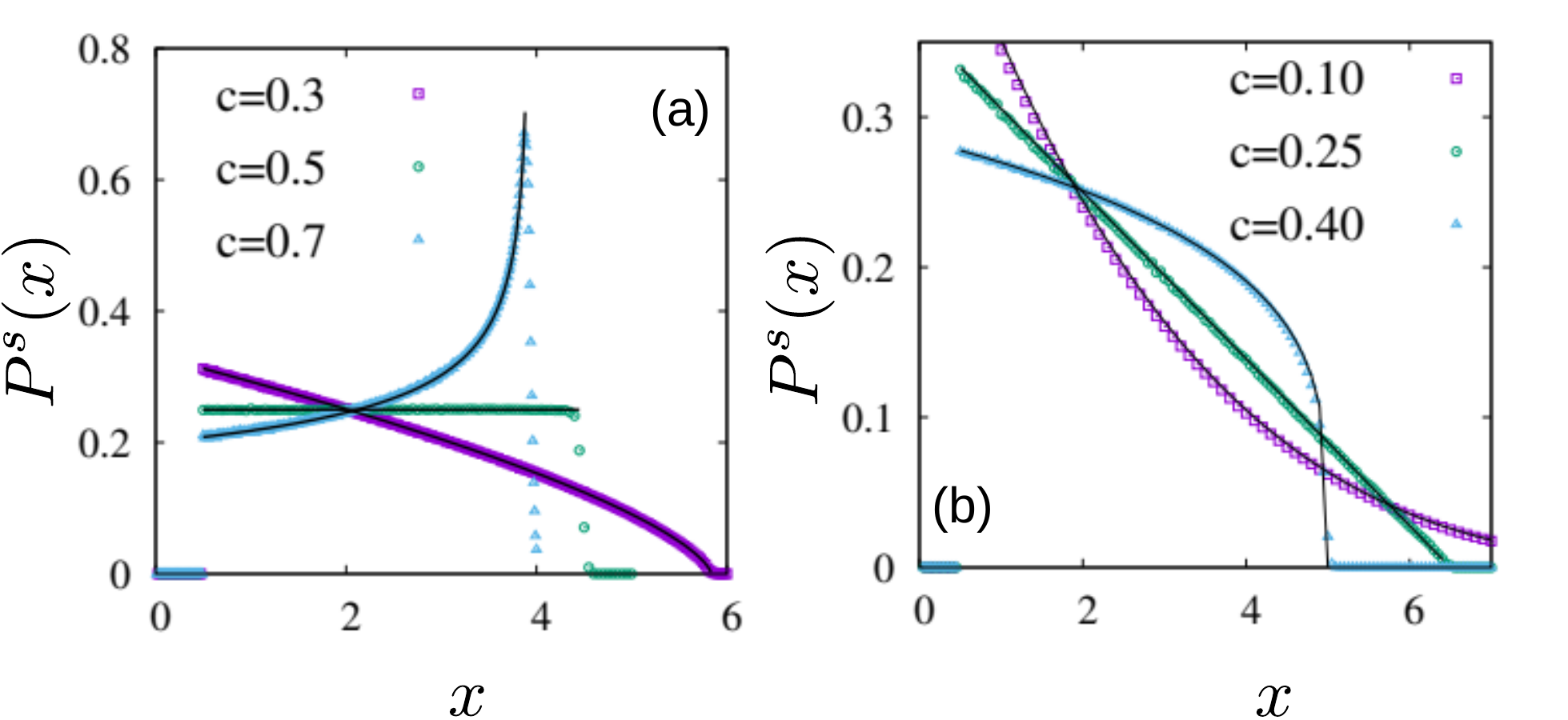}
\caption{Comparison of the stationary wealth distribution obtained in Eq.~\eqref{e:ordereddist} [solid lines] for the homogeneous case with numerical simulation [symbols] of $2000$ agents with $v=1$, $r=0.5$ and $x_r=0.5$. Panel (a) shows the transition of the stationary distribution from a decreasing to an increasing nature as $c$ increases ($\gamma=r/c$ decreases). Panel (b) shows the second transition in the region $r>c$, where the distribution changes from a decreasing concave up to a decreasing concave down shape at $c=r/2$.}
\label{f:ordered}
\end{figure}
Note that, $v_{\text{mf}}$ is positive for $x<x^*=(v+c\bar{x})/c$ and negative for $x>x^*$. Thus we expect $P^s(x)$ to vanish for $x>x^*$ as there is no current in the stationary state, which pushes a process beyond $x^*$. The stationary state distribution thus has a finite support in $x\in[x_r,x^*].$ This is also indicative from the wealth trajectory of a single agent [see Fig.~\ref{f:traj} (b)]---in between two resetting events, the growth of the wealth decreases progressively with time or equivalently at larger wealth values. The distribution in Eq.~\eqref{eq:ss} is normalized correctly. Now, using the Eq.~\eqref{eq:selfconsist} and Eq.~\eqref{f:ordered} $\bar{x}$ can be self-consistently calculated,
\begin{align}
\bar{x}=x_r+\frac{v}{r}.
\end{align}
Putting this back in Eq.~\eqref{eq:ss}, we get,
\begin{align}
P^s(x)=\frac{r\left[v\left(1+\frac{c}{r}\right)-c(x-x_r)\right]^{r/c-1}}{[v\left(1+\frac{c}{r}\right)]^{r/c}}\Theta(x^*-x).\label{e:ordereddist}
\end{align}
This shows interesting shape transitions for different values of $r/c$. We first note that the exponent on the rhs changes sign at $r=c$, either side of which shows very different qualitative behavior for the stationary distribution, as discussed below.

\textbf{\textit{Small taxation $c<r$:}} Here using Eq.~\eqref{e:ordereddist}, $P^s(x)$ peaks at $x=x_r$ decays like a power-law at large $x$, becoming zero at $x=x^*$. The peak near $x=x_r$ and the subsequent decay suggests that, poor agents (those with less wealth) are more likely in the society. Interestingly, there is another transition characterized by a concave up to a concave down shape change of the stationary distribution at $r/c=2$ (as shown in Fig.~\ref{f:ordered} (b)). This transition, occuring due to the change in sign of the second derivative at $c=r/2$, is already a sign of what happens at higher taxation, which we will see soon. Note that, for $r\gg c$, one recovers the non-interacting result (Eq.~\eqref{nonint}) for $P^s(x)$.

\textbf{\textit{High taxation $c>r$:}} Using Eq.~\eqref{e:ordereddist}, we see that $P^s(x)$ increases with increase in $x$ and diverges at the upper boundary $x=x^*=x_r+v\left(1/c+1/r\right)$. This indicates that, the richer agents are more likely in this parameter regime. However, it is noteworthy that the right boundary itself decreases with increase in $c$. In fact, the upper bound becomes close to the average stationary wealth, $x^*\to\bar{x}$, as $c\to\infty$ and the distribution tends to a $\delta$-function, $P^s(x)\to\delta(x-\bar{x})\Theta(\bar{x}-x)$.

\textbf{\textit{Critical taxation $c=r$:}} The transition between the above mentioned contrasting behaviors occur at this point. It can be seen from Eq.~\eqref{e:ordereddist} that the stationary wealth distribution $P^s(x)$ becomes flat in $x\in[x_r,x^*]$ with value, $P^s(x)=r/(2v)$, indicating that an agent is equally likely to be rich or poor.

The transitions in the shape of $P^s(x)$, as predicted above, is compared with numerical simulation in Fig.~\ref{f:ordered}, which shows excellent match. Note that the shape of the steady state distribution does not change with the growth rate $v$ qualitatively; it only depends on the competition between the resetting rate and the collection parameter. However, for fixed $r$ and $c$, the range of $P^s(x)$ increases with the increase in $v$, since $x^*$ increases with increase in $v$.

\section{Inhomogeneous systems}
Till now, we have considered the system to be homogeneous, i.e., all the parameters {\it viz.} growth rate, collection, and distribution are the same for each agent. Such systems are, of course, never found in nature. So, we proceed to inhomogeneous systems, where the parameters of the different agents are not the same but chosen from a distribution. The distribution of each agent is different, but the qualitative features remain the same as seen in the previous situation. The average stationary wealth of the $i$th agent gives a good idea of how inhomogeneity affects the economy. To this end, in the following, we calculate the distribution using the modified mean-field drift and then obtain the average stationary wealth self-consistently.

\begin{figure}
\includegraphics[width=\hsize]{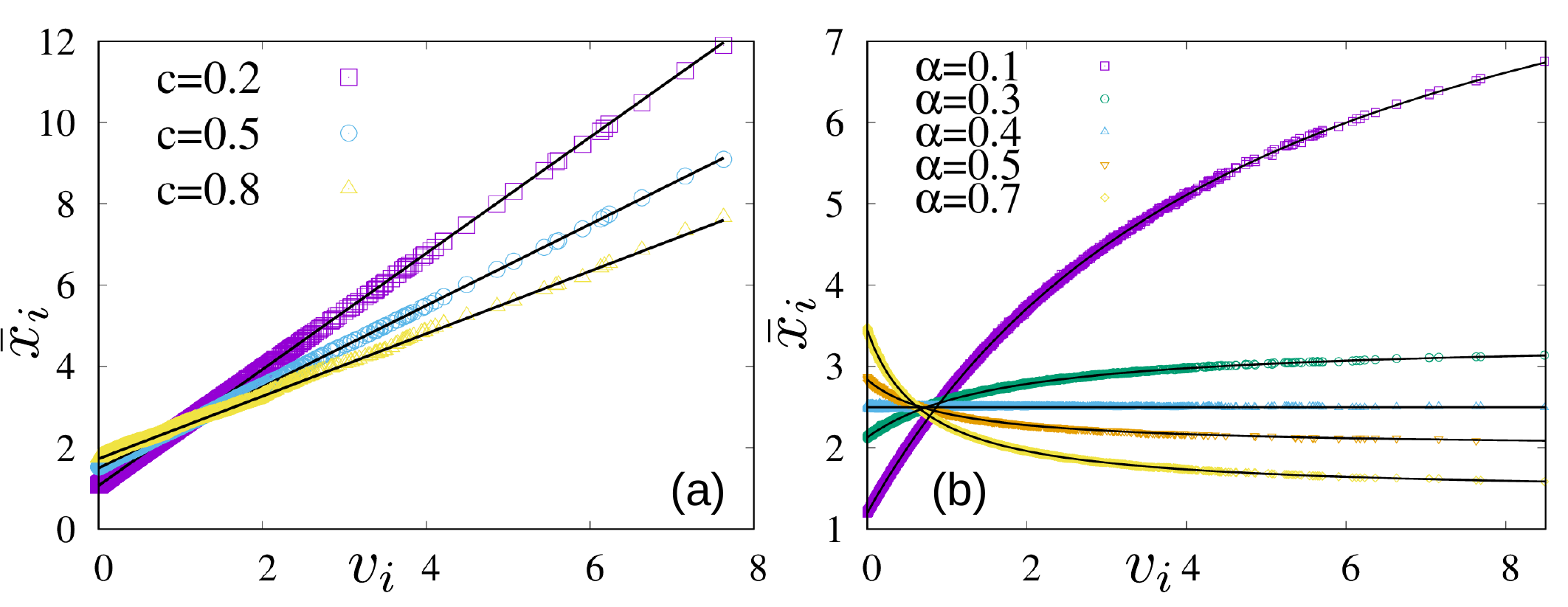}
\caption{Average wealth for exponential growth rate distribution for $r=0.5$, $x_r=0.5$, $v_0=1$ and $N=2000$: (a) compares the analytical prediction [solid lines] in Eq.~\eqref{diff_v} for uniform collection with numerical simulation [symbols] where $\bar{x}_i$ increases linearly with $v_i$. (b) compares the analytical prediction [solid lines] in Eq.~\eqref{exp:xbar_diffcv} with numerical simulations [symbols] when the collection parameter of the agents are  proportional to their respective growth rates.}
\label{f:expon1}
\end{figure}

First, let us consider that only the growth rate of each agent is different. This is a good place to start from, since in a real world, the incomes of the different people in the society are different~\cite{ununiformwealth1,ununiformwealth2}. One of the simplest, yet observable income distribution is the exponential distribution~\cite{incomedist1,exponential1}, given by $G(v)=v_0^{-1}\exp\left(-v/v_0\right)$. For such a distribution majority of the agents have a small growth rate and the number of agents with larger growth rate decay exponentially. Our ergodicity assumption still holds as $N\to\infty$. Using the fact that the redistribution parameter is still same $d_i=N^{-1},\,\forall\,i$) we can write,
\begin{align}
R_s=N^{-1}c\sum_{i}\bar{x}_i,
 \label{eq:rs}
 \end{align}
where
$ \bar{x}_i=\int_0^{\infty}dx_i\,x_i\, P^s_i(x_i)$. The Fokker-Planck equation for each agent is modified to,
\begin{align}
 \frac{\partial}{\partial x_i}[v_{\text{mf}}(x)P^s_i(x_i)]+r P^s(x_i)=r\delta(x_i-x_r),
 \label{mf:FPdisorder}
 \end{align}
with $v_{\text{mf}}=(v_i+R_s-c x_i)$. The solution of this again can be easily obtained,
 \begin{align}
P^s_i(x_i)=\frac{r}{(v_i+R_s-cx_r)^{r/c}}(v_i+R_s-cx_i)^{r/c-1}.
\end{align}
The above expression, along with the self-consistency relation, leads to, 
\begin{align}
\bar{x}_i=\int_0^{\infty}dx_i\,x_i\, P^s_i(x_i)=\frac{v_i+R_s+rx_r}{c+r}.
\label{eq:x_i}
\end{align}
Again, using this in the relation $R_s=N^{-1}c\sum_i\bar{x}_i$ and solving for $R_s$, we get,
\begin{align}
R_s=\frac{c}{r}\sum_i \frac{v_i}{N}+c x_r.
\label{eq:selfconst3}
\end{align}
For large $N$, we can replace the sum in Eq.~\eqref{eq:selfconst3} by an integral
\begin{align}
R_s=\frac{c}{r}\int_0^{\infty} dv\, v\, G(v)+c x_r=\frac{c v_0}{r}+cx_r.
\end{align}
Putting this back in Eq.~\eqref{eq:x_i}, the average wealth of the $i$th agent,
\begin{align}
\bar{x}_i=\frac{1}{c+r}\left[v_i+(r+c) x_r+\frac{cv_0}{r}\right].
\label{diff_v}
\end{align}
Thus, the average wealth of the agents increases linearly with its respective growth rate implying that the rich would always stay richer. Comparison of this analytical prediction with numerical simulations in Fig.~\ref{f:expon1} (a) shows excellent agreement.

Now, let us consider that the collection parameter of each agent is proportional to its growth rate, $c_i=\alpha v_i$ with the constant of proportionality $\alpha>0$. This is reasonable from the socio-economic point of view where a person earning more,  pays more tax than someone with a low income.
The solution of the Fokker-Planck equation for each agent [same as Eq.~\eqref{mf:FPdisorder} with $c\,$ replaced by $c_i$] is thus,
\begin{align}
P^s_i(x)=\frac{r}{(v_i+R_s-c_i x_r)^{r/c_i}}(v_i+R_s-c_ix)^{r/c_i-1}.
\end{align}
This leads to,
\begin{align}
\bar{x}_i=\int_0^{\infty}dx\,x\, P^s_i(x)=\frac{v_i+R_s+rx_r}{c+r}.
\label{eq:x_idiffcv}
\end{align}
Now, using, Eq.~\eqref{eq:x_idiffcv} in a modified form (with $c$ replaced by $c_i$) of Eq.~\eqref{eq:rs} ,
\begin{align}
R_s=\frac{1}{N}\left[\sum_{i}\frac{c_i v_i}{c_i+r}+(R_s+x_r)\sum_{i}\frac{c_i}{c_i+r}\right].
\end{align}
In the large N limit, the sums can again be converted to integrals and using the form of $c_i=\alpha v_i$, we have for $R_s$,
\begin{align}
R_s=\frac{\alpha(I_1+rx_rI_2)}{1-\alpha I_2},
\label{Rs_diffcv}
\end{align}
where 
\begin{align}
I_1=\int_0^{\infty} dv\frac{v^2\, G(v)}{\alpha v+r}\quad{\&}\quad I_2=\int_0^{\infty} dv\frac{v\, G(v)}{\alpha v+r}.
\end{align}
%=\frac{\alpha v_0-r}{\alpha ^2}-\frac{r^2 e^{\frac{r}{\alpha  v_0}} \text{E}_1\left(-\frac{r}{\alpha  v_0}\right)}{\alpha^3 v_0}
%and
%\begin{align}
%I_2&=\int dv\frac{v G(v)}{\alpha v+r}=\frac{1}{\alpha}+\frac{r e^{\frac{r}{\alpha  v_0}} \text{E}_1\left(-\frac{r}{\alpha  v_0}\right)}{\alpha^2 v_0}
%\end{align}
 Using Eq.~\eqref{Rs_diffcv} and the values of $I_1$ and $I_2$ (see Supplemental material~\cite{SM}), 
\begin{align}
\bar{x}_i=\frac{1}{\alpha }+\frac{v_0 e^{-\frac{r}{\alpha  v_0}} \left(r-\alpha v_0-\alpha r x_r\right)}{r (r+\alpha  v_i) \text{E}_1\left(-\frac{r}{\alpha  v_0}\right)},
\label{exp:xbar_diffcv}
\end{align}
where E$_n$ denotes exponential integral $\text{E}_n (z)=-\int_{-z}^{\infty } \frac{e^{-t}}{t^n} \, dt$. The right hand side of the above equation is always positive and shows many interesting features. 
\begin{figure*}
\centering
\includegraphics[width=\hsize]{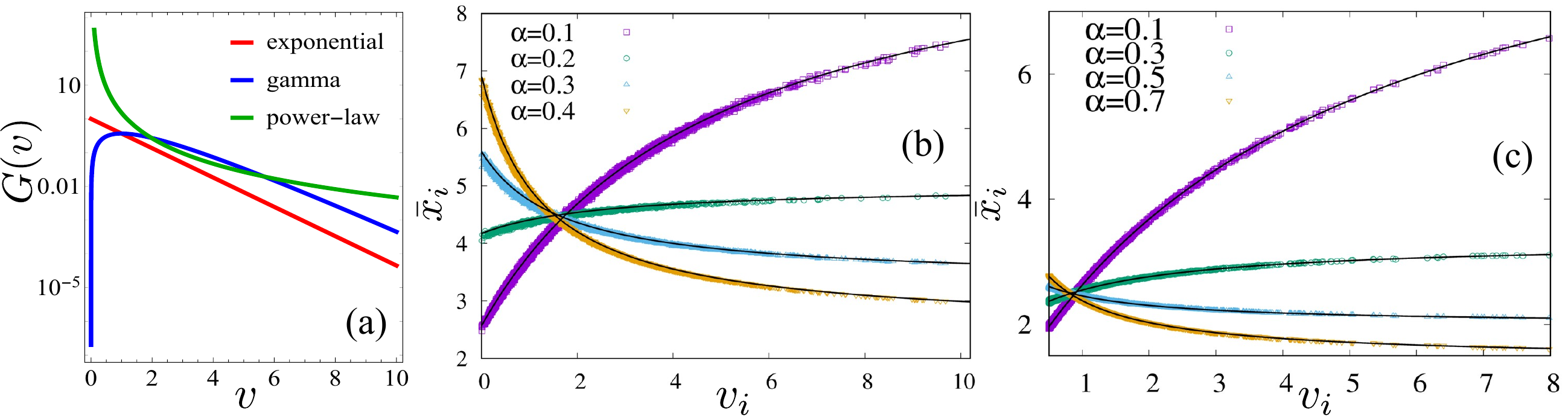}
\caption{(a) shows the three different growth rate distributions studied in this paper. Both (b) and (c) compare the average stationary wealth when the tax collection from each agent is proportional to its respective growth rate. In particular, (b) compares the average wealth when the growth rates of the agents are chosen from a Gamma distribution (with $k=1$, $v_0=1$) as predicted in Eq.~\eqref{xbar-gamma} [solid lines] with numerical simulations [symbols]. (c) compares the average wealth when the growth rates of the agents are chosen from a power-law  distribution (with $a=2$, $v_0=0.1$) as predicted in Eq.~\eqref{xbar-plaw} [solid lines] with numerical simulations [symbols]. In both (b) and (c), $r=0.5$, $x_r=0.5$ and $N=2000$.}
\label{f:otherdist}
\end{figure*}
The average stationary wealth $\bar{x}_i$ shows interesting behavior on either side of a critical value of the collection rate $\alpha=\alpha^*$, which is obtained by $\frac{\partial \bar{x}_i}{\partial v_i}\big|_{\alpha=\alpha^*}=0$. Using Eq.~\eqref{exp:xbar_diffcv}, we get $\alpha^*=\frac{r}{v_0+r x_r}$. 

\textbf{ \textit {Case $\bm{\alpha<\alpha^*}$:}} Here the second term on the rhs of Eq.~\eqref{exp:xbar_diffcv} is negative and goes to zero $\sim(r+\alpha v_i)^{-1}$. Thus, $\bar{x}_i$s increase with increase in $v_i$s, albeit slower than the uniform collection rate case, and saturates to a maximum value of $\alpha^{-1}$.
 Thus though agents with higher growth rates have a larger average wealth, there is a finite upper bound on the same.
  
\textbf{ \textit { Case $\bm{\alpha<\alpha^*}$:}} In this case, the  second term on the rhs of Eq.~\eqref{exp:xbar_diffcv} is positive and decays to $0$ as $\sim(r+\alpha v_i)^{-1}$. Thus, here, $\bar{x}_i$s decrease with increase in $v_i$s and decays to a minimum value of  $\alpha^{-1}$, implying that agents with small growth rates have more wealth than the ones with larger growth rates.

\textbf{ \textit { Case $\bm{\alpha=\alpha^*}$:}} At this point, the second term on the rhs of Eq.~\eqref{exp:xbar_diffcv} is always zero independent of the value of $v_i$. Thus, $\bar{x}_i$ remains constant $1/\alpha$ for all agents, irrespective of its individual growth rate. This indicates an ideal economic equality.
  
The above theoretical predictions are compared with numerical simulations in Fig.~\ref{f:expon1} (b) and show excellent agreement. 

 Thus, we see that in systems with inhomogeneous growth rates and homogeneous taxation,  the average wealth of an agent is proportional to its growth rate Eq.~\eqref{diff_v}, implying a huge disparity in the average wealth of individuals with small and large growth rates. This disparity can be easily controlled by making individual taxations proportional to the respective growth rates. We find an optimal value of the proportionality constant $\alpha=\alpha^*$, for which $\bar{x}_i$ becomes completely flat, indicating an economic equality. Increasing taxation beyond this, however, results in a reverse disparity, where agents with lesser growth rates have higher $\bar{x}_i$.
In fact, these features are quite robust, and the qualitative picture remains the same for other growth rate distributions as well. To demonstrate this, we consider two more physically relevant income distributions, namely a gamma-distribution and a power-law distribution, and show that the qualitative features seen in the exponential income distribution with proportional taxation are the same in both cases.

 {\it Gamma distribution} The growth rates are chosen from an income distribution, which in this case, is given by $G(v)=v_0^{-k}v^{k-1}\exp(-v/v_0)/\Gamma(k)$. Unlike the exponential distribution which is peaked about zero, the gamma-distribution vanishes close to zero, peaks about some intermediate value and then decays exponentially [see Fig.~\ref{f:otherdist} (a)]. Gamma-distribution is a two parameter distribution which has been a convenient descriptive model of income distribution in the economics literature~\cite{gammadistincome,gammadistincome2}, where the parameters $v_0$ and $k$ are obtained by fitting with real world data and vary country-wise. The average wealth of an agent in a society where the income follows a gamma distribution is given by~\cite{SM},
 \begin{align}
 \bar{x}_i=\frac{1}{\alpha}+\frac{r-\alpha  k v_0-\alpha  r x_r}{\alpha(r+\alpha  v_i) \left(k e^{\frac{r}{\alpha  v_0}} E_{k+1}\left(\frac{r}{\alpha  v_0}\right)-1\right)}.
 \label{xbar-gamma}
 \end{align}
 {\it Power-law distribution} The income distribution in this case is given by $G(v)=a v_0^{a}v^{-(a+1)}$ with $a>1$ for $v>v_0$. Power-law distribution is also a well-used income distribution in economics literature~\cite{powerlawdist1,powerlawdist1.1,powerlawdist2,powerlawdist3}. This is again a two-parameter distribution model, where the parameters $v_0$ and $a$ are obtained by fitting with the real world data. Compared to the two previously considered distributions, the power-law tails have flatter tails [see Fig.~\ref{f:otherdist} (a)]. The average wealth of an agent in a society with a power-law distributed income can be computed~\cite{SM} to yield the explicit form,
\begin{align}
\bar{x}_i&=\frac{v_i}{r+\alpha  v_i}+
\frac{\left(rx_r +\frac{a v_0}{a-1}{}_2F_1\left(1,a-1;a;-\frac{r}{\alpha  v_0}\right) \right)}{ \left(r+\alpha  v_i\right) \left(1-\,{}_2F_1\left(1,a;a+1;-\frac{r}{\alpha  v_0}\right)\right)},
\label{xbar-plaw}
\end{align}  
where ${}_2F_1(a,b;c;z)$ denotes the Gauss hypergeometric function. 

In both of the above cases, there is an optimal $\alpha=\alpha^*$ for which the society exhibits economic equality (average wealth is constant for any value of the growth rate). Below $\alpha^*$, the average wealth ($\bar{x}_i$) increases with the increase in respective growth rate ($v_i$) and saturates to a value $\alpha^{-1}$ as $v_i\to\infty$. Above $\alpha^*$, one can again see the reverse disparity, where agents with lesser growth rates tend to have higher average wealth. Thus, the qualitative features remain the same compared with the exponential case. The theoretical predictions for the Gamma-distribution in Eqs.~\eqref{xbar-gamma} (for $k=1$) and power-law distribution \eqref{xbar-plaw} (for $a=2$) are compared with numerical simulations in Fig.~\ref{f:otherdist} (b) and (c) respectively, and show excellent match.  
\section{Summary and Conclusion}
 In summary, we study a model of $N$ agents, whose wealth grows at a constant rate, undergoes resetting events at a constant rate, and interacts via a tax-dynamics---all agents donate a part of their wealth, which is in turn redistributed among themselves equally. This simple model mimics some of the fundamental attributes of a society's economy. In the absence of taxation, the wealth distribution reaches an exponentially decaying stationary state. The introduction of the tax dynamics results in some very interesting stationary properties. To treat the effect of taxation, which can also be looked at as an $N$-particle interaction, we make an ergodicity hypothesis in the $N\to\infty$ limit. This enables us to decouple the wealth evaluation equation of a single agent in terms of a mean-field drift $v_{\text{mf}}$ in the stationary state. Using this, we first show that for a homogeneous system, taxation not only introduces a finite upper cut-off,  the wealth distribution also shows interesting shape transitions with the taxation strength. The distribution shows a decreasing curve for low taxation, implying an abundance of agents with lesser wealth. This changes to an increasing curve as the taxation is increased, indicating that if the taxation is high, agents are more likely to be rich.  We explicitly calculate the critical line $r/c=1$ for this transition, which is also verified from numerical simulations. We then look at inhomogeneous systems, where the growth rates are chosen from different distributions seen in real world scenarios. We find that in the case of uniform tax collection, the average wealth of the agents increases linearly with their respective growth rates. However, if the collection rates are proportional to the respective growth rates ($c_i=\alpha v_i$), then we see a host of interesting features: Below a critical value $\alpha^*$, the average wealth increases with an increase in $v_i$ and saturates to a value $\alpha^{-1}$, while above $\alpha^*$ there is a reverse disparity, where the average wealth of the agents with higher growth rate is lesser. The transition between the two above-mentioned contrasting behaviors occurs at the critical point $\alpha=\alpha^*$, where $\bar{x}_i$ is a constant, independent of the individual growth rate $v_i$. Thus, within the limits of the model, we get an optimal collection rate, which produces complete economic equality in the society.
 
 Though our model is very simple and lacks the many complexities of real world economies, it has some fundamental aspects of the same. The stationary state analysis in our work yields nice closed-form analytical results, predicting the exact transition lines and optimal parameter values in the different cases. This model can be generalized to include more complexities systematically, e.g., an additive noise at each time step---indicating small-time fluctuations in wealth, wealth dependent resetting value for different agents,  presence of an underlying income dynamics like~\cite{srgbm}. The analytical formalism would work in the $N\to\infty$ limit, as long as the system reaches a well-defined stationary state. Another future direction is the study of extreme value statistics, which is very important for economies~\cite{extremevalueeco1,extremevalueeco2}. It would be interesting to see if, in the spirit of \cite{Singh2021,extremevaluereset2} one can study the extreme value statistics of this model.

%%%%%%%%%%%%%%%%%%%%%%%%%%%%%%%%%%%%%%%%%%%%
\acknowledgments
The author thanks Urna Basu and Chitrak Bhadra for useful discussions during the course of the work.
%%%%%%%%%%%%%%%%%%%%%%%%%%%%%%%%%%%%%%%%%%%%

\end{document}

\maketitle

\section{Calculation of average stationary wealth for exponentially distributed growth rates}
In this section we show the detailed computation of the average wealth of an agent in a society, where the individual growth rates (incomes) are chosen from an exponential distribution,
\begin{align}
G(v)=\frac{1}{v_0}\exp\left(-v/v_0\right).
\label{expdist}
\end{align}
and the tax collection rate is proportional to the individual wealth of an individual. The average wealth of an agent, given by Eq.~(25) in the main text,
\begin{align}
\bar{x}_i=&\frac{v_i+R_s+r x_r}{c+r}
\label{barxi}
\end{align}
with,
\begin{align}
R_s=\frac{\alpha(I_1+rx_rI_2)}{1-\alpha I_2}.
\label{rs}
\end{align}
The functions $I_1$ and $I_2$ in the above equation are given by,
\begin{align}
I_1&=\int_{0}^{\infty}dv\frac{v^2 G(v)}{\alpha v+r}\label{i1:app}\\
\text{and}\qquad I_2&=\int_{0}^{\infty}dv\frac{v G(v)}{\alpha v+r}
\label{i2:app}
\end{align}
Using Eq.~\eqref{expdist} in \eqref{i1:app}, we get,
\begin{align}
I_1&=\frac{\alpha v_0-r}{\alpha ^2}-\frac{r^2 e^{\frac{r}{\alpha  v_0}} \text{E}_1\left(-\frac{r}{\alpha  v_0}\right)}{\alpha^3 v_0}
\end{align}
and using Eq.~\eqref{expdist} in \eqref{i2:app}, we get,
\begin{align}
I_2&=\frac{1}{\alpha}+\frac{r e^{\frac{r}{\alpha  v_0}} \text{E}_1\left(-\frac{r}{\alpha  v_0}\right)}{\alpha^2 v_0}
\end{align}
E$_n$ denotes exponential integral $\text{E}_n (z)=-\int_{-z}^{\infty } e^{-t}\,t^{-n} \, dt$. 
Finally, substituting the values of $I_1$ and $I_2$ in Eq.~\eqref{rs} and using Eq.~\eqref{barxi}, 
\begin{align}
\bar{x}_i=\frac{1}{\alpha }+\frac{v_0 e^{-\frac{r}{\alpha  v_0}} \left(r-\alpha v_0-\alpha r x_r\right)}{r (r+\alpha  v_i) \text{E}_1\left(-\frac{r}{\alpha  v_0}\right)}
\label{xbar_diffcv}
\end{align}
This is the result given in the main text.
\section{Gamma distributed growth rates}
In this section we consider the growth rates to be drawn from a Gamma distribution given by,
\begin{align}
G(v)=\frac{v^{k-1}}{v_0^{k}}\frac{\exp(-v/v_0)}{\Gamma(k)}
\label{gammadist}
\end{align}
Using Eq.~\eqref{gammadist} in Eq.~\eqref{i1:app},
\begin{align}
I_1&=\frac{k (k+1) v_0 e^{\frac{r}{\alpha  v_0}} \text {E}_{k+2}\left(\frac{r}{\alpha  v_0}\right)}{\alpha }\\
I_2&=\frac{k e^{\frac{r}{\alpha  v_0}} \text{E}_{k+1}\left(\frac{r}{\alpha  v_0}\right)}{\alpha }
\end{align}
Now, we can use the obtained values of $I_1$ and $I_2$ in Eq.~\eqref{barxi} to get,
\begin{align}
\bar{x}_i=\frac{1}{\alpha}+\frac{r-\alpha  k v_0-\alpha  r x_r}{\alpha(r+\alpha  v_i) \left(k e^{\frac{r}{\alpha  v_0}} E_{k+1}\left(\frac{r}{\alpha  v_0}\right)-1\right)}
\label{gammabarx}
\end{align}
The value of $\alpha=\alpha^*$, for which there is complete equality can be obtained by $\frac{\partial \bar{x}_i}{\partial v_i}\big|_{\alpha=\alpha^*}=0$,
\begin{align}
\alpha^*=\frac{r}{kv_0 +r x_r}.
\label{alpha*barx}
\end{align}
Note that, the exponential distribution is a special case of Gamma distribution (k=1), and putting $k=1$ in Eqs.~\eqref{gammabarx} and \eqref{alpha*barx} we get back the results obtained for the exponential distribution.
\section{Power-law distribution}
In this section we consider the growth rates to be drawn from a power-law given by,
\begin{align}
G(v)=a v_0^{a}v^{-(a+1)}\quad\text{with }a>1
\label{plawdist}
\end{align}
Note that the power-law distribution defined above has a lower cut-off at $v=v_0$, thus the integrals in Eqs.~\eqref{i1:app} and \eqref{i2:app} are modified to,
\begin{align}
I_1&=\int _{v_0}^{\infty}dv\frac{v^2 G(v)}{\alpha v+r}\\
\text{and}\qquad I_2&=\int _{v_0}^{\infty}dv\frac{v G(v)}{\alpha v+r}
\end{align}
Using the above equations along with Eq.~\eqref{plawdist}, we get,
\begin{align}
I_1&=\frac{a v_0}{a-1}\frac{  \, _2 F_1\left(1,a-1;a;-\frac{r}{\alpha  v_0}\right)}{\alpha  }\\
I_2&=\frac{\, _2F_1\left(1,a;a+1;-\frac{r}{\alpha  v_0}\right)}{\alpha }
\end{align}
Finally, using Eq.~\eqref{barxi}, we find,
\begin{align}
\bar{x}_i&=\frac{v_i}{r+\alpha  v_i}+
\frac{\left(rx_r +\frac{a v_0}{a-1}{}_2F_1\left(1,a-1;a;-\frac{r}{\alpha  v_0}\right) \right)}{ \left(r+\alpha  v_i\right) \left(1-\, _2F_1\left(1,a;a+1;-\frac{r}{\alpha  v_0}\right)\right)}
\end{align}
The presence of the Hypergeometric functions in the final expression for $\bar{x}_i$ makes it difficult to obtain a closed-form expression for $\alpha^*$. But we illustrate this graphically in the following.
Using, $\frac{\partial \bar{x}_i}{\partial v_i}\big|_{\alpha=\alpha^*}=0$, we get a transcendental equation, $g_1(\alpha,v_i)=g_2(\alpha,v_i)$, where
\begin{align}
g_1(\alpha,v_i)&=\frac{r}{(r+\alpha v_i)^2}\\
g_2(\alpha,v_i)&=\frac{\alpha}{(r+\alpha v_i)^2}  \frac{\left(rx_r +\frac{a v_0}{a-1}2F_1\left(1,a-1;a;-\frac{r}{\alpha  v_0}\right) \right)}{ \left(1-\, _2F_1\left(1,a;a+1;-\frac{r}{\alpha  v_0}\right)\right)} 
\end{align}
\begin{figure}
\centering\includegraphics[width=0.55\hsize]{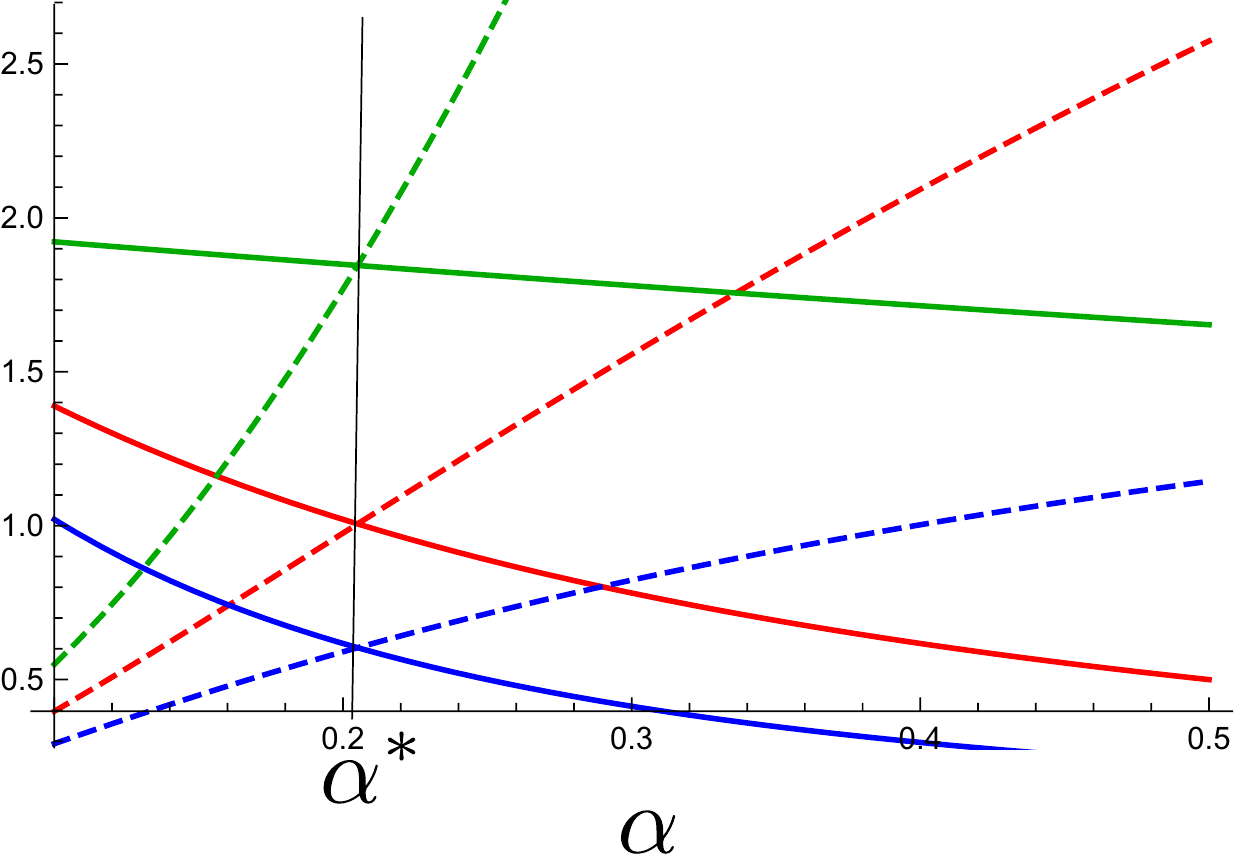}
\caption{Plot illustrating the existence of optimal $\alpha=\alpha^*$ for power-law distributed growth rates by graphical solution of $\frac{\partial \bar{x}_i}{\partial v_i}\big|_{\alpha=\alpha^*}=0$. The solid and dashed lines indicate $g_1(\alpha,v_i)$ and $g_2(\alpha,v_i)$ for $v_i=0.1$ (green), $1.0$ (red) and $2.0$ (blue) for $r=0.5,\,x_r=0.5,\,v_0=1.0,\,a=2$. The locus of the intersection of the solid and dashed curves is $\alpha=\alpha^*$.}
\label{f:plaw}
\end{figure}
Figure~\ref{f:plaw} shows a plot of $g_1(\alpha,v_i)$ [in solid lines] and $g_2(\alpha,v_i)$ [in dashed lines] against $\alpha$ for different values of $v_i$. The locus of the intersection of the solid and dashed lines yields yields the straight line $\alpha=\alpha^*$.